# An Automated Power Conservation System (APCS) using Particle Photon and Smartphone

**Chandra Sekhar Sanaboina[1]\*, Harish Bommidi[2]**

[1]Department of Computer Science and Engineering, University College of Engineering, JNTUK – Kakinada, India
[2]Department of Computer Science and Engineering, University College of Engineering, JNTUK – Kakinada, India

*Corresponding Author: chandrasekhar.s@jntucek.ac.in   Tel.: ++919885258544*

*Abstract*— Nowadays, people use electricity in all aspects of their lives so that electricity consumption increases gradually. There can be wastage of electricity due to various reasons, such as human negligence, daylighting, etc. Hence, conservation of energy is the need of the day. This paper deals with the fabrication of an "Automated Power Conservation System (APCS)" that has multiple benefits like saving on power consumption there by saving on electricity bills of the organization, eliminating human involvement and manpower which is often required to manually toggle the lights and electrical devices on/off, and last but most importantly conserve the precious natural resources by reducing electrical energy consumption. Two IR sensors are used in this project and these two sensors are used for detecting the presence of a person in the classroom. When the existence of the person is detected by the APCS it automatically turns on the fans and lights in that classroom and during the absence they will be automatically turned off, thus paving the easiest way to conserve power. This hardware is integrated with the Android app, where the user can get data on his smartphone regarding the number of fans and lights that are turned on at a particular instance of time. The user can also switch on/off the fans and lights from anywhere in the world by using the Android App.

### I. INTRODUCTION

An Automated Power Conservation System (APCS) is a system, which monitors and operates the electrical appliances in accordance to the presence of a person in the classroom and it can be accessed from a remote place through internet and an Android App (UI). (i.e,. APCS turns on/off the fans and lights in the classroom based on the presence of the people in the classroom). APCS also maintains the count of the number of persons present in the classroom and the data obtained is pushed on to the cloud. It can also generate the report and give the status of electrical appliances (on/off) on a real-time basis.

APCS consists of a microcontroller for monitoring and coordinating with sensors and appliances of a room. Here the Particle Photon Chip is used as a microcontroller. Particle Photon is a microcontroller with inbuilt Broadcom Wi-Fi device which assists the photon to connect to the internet. The particle is connected with two IR sensors (INPUTS) and Relay (OUTPUTS). A Web IDE (Integrated Development Environment) is used to program the particle photon. The program specifies a task to particle photon based on inputs read from the two sensors and uploads the data into Cloud. The admin can monitor these data from a remote place by using his/her smartphone via the internet. Android studio is used for developing Android App (UI). The IR sensors used for detecting person entry/exit status and relay module are used for turn ON/OFF the electrical appliances. Moreover, the user is also facilitated to operate the appliances remotely.

APCS saves unwanted power consumption of the University or any other organization, eliminating human involvement and manpower which is often required to manually switch on/off the lights and the electrical device.

The rest of the paper is structured as follows. Section II explains Literature survey of the IoT and Automation. Section III elucidates the proposed model along with the description of all the components used in the experiment. Section IV describes the system design and implementation. Section V deals with the algorithm and flowchart for APCS. Section VI shows the Experimental setup. Section VII gives the experimental results. Finally, Section VIII discusses the conclusion and future work.

### II. LITERATURE SURVEY OF IoT AND AUTOMATION

The Internet of Things (IoT) conceptually embodies intelligent visions of automating the day to day activities [1]. Ideally, loT will optimize our future routines with intelligent and robust systems that will make our life not only easy but also fast based upon our preferences and priorities like morning alarms, coffee timing, medicine uptake etc. Its vast applications will make our travel arrangements intelligently, by giving frequent updates and weather data. In short, loT has the power to meet our every need before we even need to realize what we want and will need. Interconnectedness and automation are the real power of loT solutions. A lot has not only made our lives easier but also has lots of potentials to drive economic value and social change [2]. But still, 85% of things still is unconnected and a security threat pervasive, for which industry has yet to conquer the real potential of loT.

Automation is a technology which enables the user to control a process or procedure with minimum human assistance[3]. Automation otherwise called as automatic control uses various control systems for operating equipment such as switching on telephone networks, processes in the factories, stabilization of ships, aircraft etc., and heat treating ovens like boilers with very minimal or condensed human intervention [4]. Some processes are partly automated but some processes have been completely automated. Automation covers applications range simple household thermostat control in air conditioning unit or thermostat controlling unit in a boiler to a large industrial control system that can control tens of thousands of input as well as output devices [5].

One of the automation wings can be a home automation system that performs the operations of various home appliances more expedient and saves energy [6]. Home automation or building automation makes life very simple nowadays and it also saves a lot of energy. It involves automatic controlling of all electrical or electronic devices in homes or even remotely through wireless communication [7]. Centralized control of security systems, lighting equipment, kitchen appliances, air conditioning units, heating devices, audio/video systems and all other equipment used in home systems is possible with this system. Automation can also be extended to the universities wherein one can automate the electrical appliances in the classrooms [8].

### III. PROPOSED MODEL AND ITS COMPONENTS

The proposed model APCS is a power-efficient Wi-Fi based intelligent automated system for a room that will monitor and control the electrical appliances without the human intervention. The model is being deployed in the classrooms for conserving the power and preventing the unwanted wastage of power. This system can work in two modes namely Automatic mode and Manual mode which is developed for our convenience. By default, the system will work with automatic mode. This mode saves power from human negligence in situations like when persons leave the room without turning off the lights and fans. This mode works on the presence of the person and person count. As soon as a person enters into the classroom the lights and fans are automatically turned on and it also keeps track of the number of persons present in the room. If the person count in the room reaches zero means it automatically turns off all the electrical appliances. In some situations, the user needs to turn on/off electrical appliances manually. To cater to these needs APCS is built to operate in a manual mode wherein the user can operate the electrical appliances in the room manually from anywhere in the world using an android app.

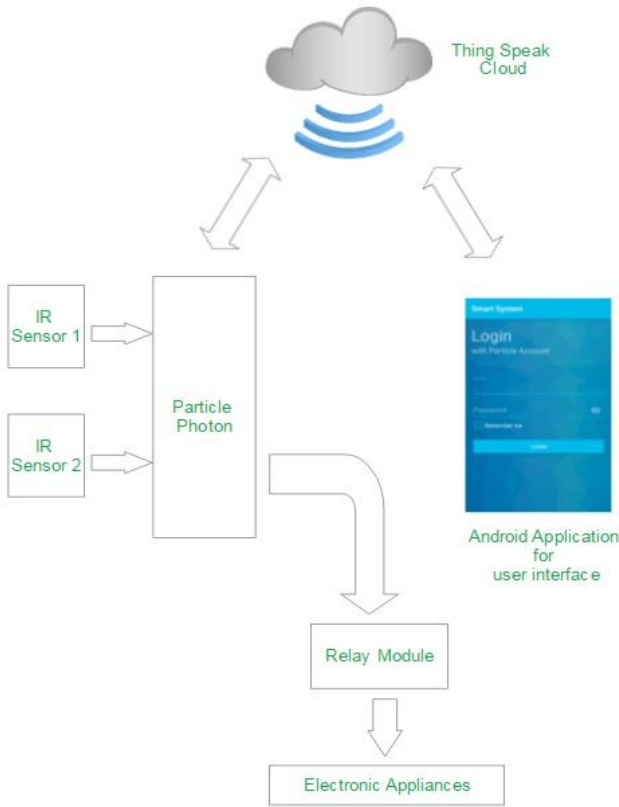

*Figure 1: An Architecture of An Automated Power Conservation System*

Figure 1 shows the Architecture of the proposed system. This system consisting of three parts mainly Hardware consisting of sensors, relay module and particle photon chip, User Interface (Android app) and Cloud which is used to store the data on a real-time basis for which Thingspeak is being used.

The most commonly known cause of energy wastage is human negligence. In most cases, humans tend to forget to turn off the electrical appliances as they left from Classroom. A smart room should be able to automatically turn off the lights and fans when it detects no person in the room. Table 1 gives a clear picture of how much power is being wasted in a month for a single room (i.e., trail room in this paper) having 4 lights and 4 fans. Ignoring the remaining factors for the wastage of power, this paper considers only one factor FOR THE WASTAGE OF POWER (I.E., HUMAN NEGLIGENCE).

Table 1: Wastage Of Power due to Human Negligence

|  | Worst Case (90%Negligence) | Average Case (50%Negligence) | Best Case (10%Negligence) |
|---|---|---|---|
| **Wastage of Power in KWH(Units)** | 163.0368 | 90.576 | 18.1152 |
| **Utilization of Power in KWH(Units)** | 18.1152 | 90.576 | 163.0368 |

The calculations are based on the theoretical power consumptions that a particular fan (60 stats) or a light (40 watts) will consume. It is not practically possible to determine how much electricity will be wasted as negligence because electricity wastage is directly proportional to human negligence. Hence, this paper considered only three cases (i.e., 90% negligence (Worst Case), 50% negligence (Average Case) and 10% negligence(Best Case)) on a trial basis and the results are tabulated as above. Each case that is considered is based on the percentage of human negligence. The table also depicts the power consumed if the appliances are in full working condition versus the power wasted due to human negligence. Practically speaking, human negligence cannot be avoided and hence there is a need to conserve power by fabricating or developing some automated tools/devices. This is the principal motto behind the development of APCS. The calculations of power consumption in our trail room which consists of 4 fans and 4 lights are elucidated in the following section. The calculations are based on the worst case negligence (i.e., 90% negligence).

**Calculation of Power Consumption of a trial room**

By definition, Power is defined as the amount of energy used per unit time and 1 watt = 1 Joule/sec

**For tube light:**

Energy consumed is 40 Joules/sec (Since tube light is 40 watts )

For 1 hour, Energy usage = 40*3600=144000J.

1 standard unit of electricity i.e. 1 kwh= 3600000Joules.

So that would be 4% of a single unit, or in other words lighting of the tube light for 25 hours would cost you 1 unit of electricity. So it would be 0.0513 kWh of single 40W tube light power usage.

For 30 days working at 8 hours per day and 90% human negligence

Power wastage = (30*8*90)/100

= 216*0.0513 (for one tube light)

= 216*0.0513*4 (for four tube lights)

= 44.3232

**For fan:**

Energy consumed is 60 Joules/sec (Since fan is 60 watts)

For 1 hour, Energy usage = 60*3600=2160000J.

1 standard unit of electricity i.e. 1 kwh= 3600000Joules.

So that would be 6% of a single unit, or in other words lighting of the fan for 24 hours would cost you 1 unit of electricity. So it would be 0.1374 kWh of single 60W fan power usage.

For 30 days working at 8 hours per day and 90% human negligence

Power wastage = (30*8*90)/100

= 216*0.1374 (for one fan)

= 216 * 0.1374 * 4 (for four fans)

= 118.7136

Hence, the total wastage of power for 4 tube lights and 4 fans is given by sum of wastage of power for 4 tube lights + wastage of power for 4 fans

Total wastage = 44.3232 + 118.7136

= 163.0368

The same value is given in Table 1 under wastage of power with 90% negligence. The remaining values given in the table are self-explanatory.

### 3.1. Hardware used in APCS

The hardware part of APCS consists of sensors, Relay module, and microcontroller. Sensors read data from the real world and send to the microcontroller. The microcontroller processes the data that is received and then controls appliances through Relay module.

A brief description of various components used in APCS are given below

#### 3.1.1 Particle Photon

Particle combines a powerful ARM Cortex M3 microcontroller with a Broadcom Wi-Fi chip in a tiny thumbnail-sized module called the PÃ˜ (P-zero). The microcontroller is the brain of particle device. It runs the program and tells hardware prototype what to do. Unlike a computer, it can only run one application (often called firmware or an embedded application). This application can be simple (just a few lines of code), or very complex (may vary from thousands to lakhs of code). The microcontroller interacts with the outside world using pins. Pins are the input and output parts of the microcontroller that are exposed on the sides of particle device. GPIO pins can be hooked to sensors or buttons to listen to the world, or they can be hooked to lights and buzzers to act upon the world. There are pins for Serial/UART communication and a pin for resetting particle device. Figure 2 shows the top view of the Particle Photon.

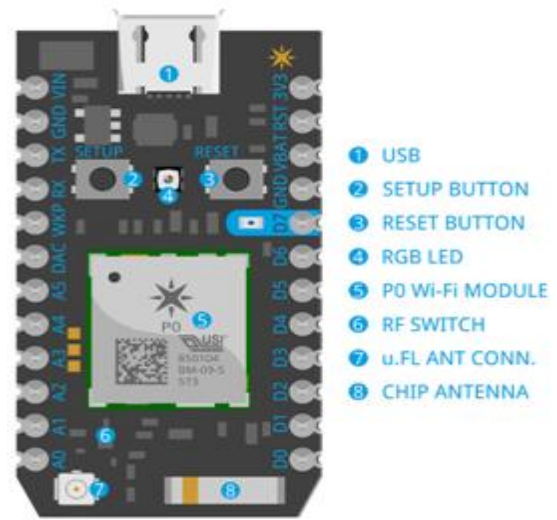

*Figure 2: Particle Photon Chip*

As shown in Figure 2 particle photon has 6 Analog I/O ports with 2 port for RX and TX, one DAC ( Digital to Analog Converter), and 8 Digital I/O port with inbuilt LED at D7 port. The particle has its own cloud where the data in our particular photon can be accessed through the Internet. Also having its own cloud, the data can be accessed by IFTTT and use it for action and trigger purpose to send an email and receive command through user easily.

Features

• Broadcom BCM43362 Wi-Fi chip

• 802.11b/g/n Wi-Fi

• 1MB flash, 128KB RAM

• On-board RGB status LED (ext. drive provided)

• 18 Mixed-signal GPIO and advanced peripherals

• Real-time operating system (FreeRTOS)

#### 3.1.2 Two-Channel Relay module

A relay is an electrically operated switch. Many relays use an electromagnet to mechanically operate a switch, but other operating principles are also used, such as solid-state relays. Relays are used where it is necessary to control a circuit by a low-power signal (with complete electrical isolation between control and controlled circuits), or where several circuits must be controlled by one signal. Since relays are switches, the terminology applied to switches applied to relays. Figure 3 shows the physical view of Two Channel Relay Module. A relay can be operated in two states:

**Normally - open (NO)** contacts connect the circuit when the relay is activated; the circuit is disconnected when the relay is inactive. It is also called a FORM A contact or make contact.

**Normally - closed (NC)** contacts disconnect the circuit when the relay is activated; the circuit is connected when the relay is inactive. It is also called FORM B contact or break contact.

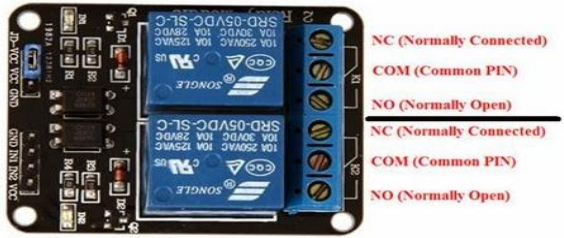

*Figure 3: 2-Channel Relay module*

**Features**

- Number of Relays: 2
- Control signal: TTL level
- Rated load: 7A/240VAC 10A/125VAC 10A/28VDC

### 3.1.3 IR Obstacle Sensor

Infrared Obstacle Sensor Module has a built-in IR transmitter and IR receiver that sends out IR energy and looks for reflected IR energy to detect the presence of any obstacle in front of the sensor module. Figure 4 shows the IR Obstacle sensor. The module has an onboard potentiometer that lets the user adjust the detection range. The sensor has a very good and stable response even in ambient light or in complete darkness. Infrared Photodiodes are different from normal photodiodes as they detect only infrared radiation. When the IR transmitter emits radiation, it reaches the object and some of the radiation reflects back to the IR receiver. Based on the intensity of the reception by the IR receiver, the output of the sensor is defined

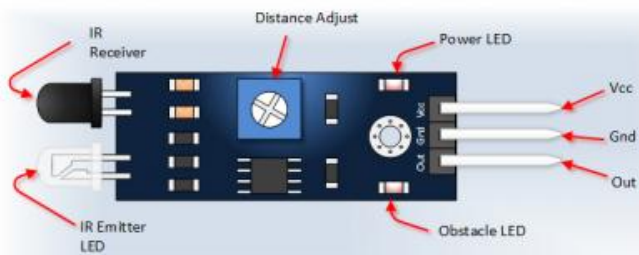

*Figure 4: IR Obstacle Sensor*

### 3.2 Mobile Application

The mobile application is developed under the Android platform. It works on all Android mobile phones which are enabled by the internet. This application requires authorization for accessing or controlling of particle device.

The mobile application fetches the data from the Thingspeak the cloud which is already uploaded by the particle photon and also sends commands to the particle device. The user interface for the application is designed in a way that enables both monitoring and control field from the device. The Mobile Application Interface is shown in Figure 5

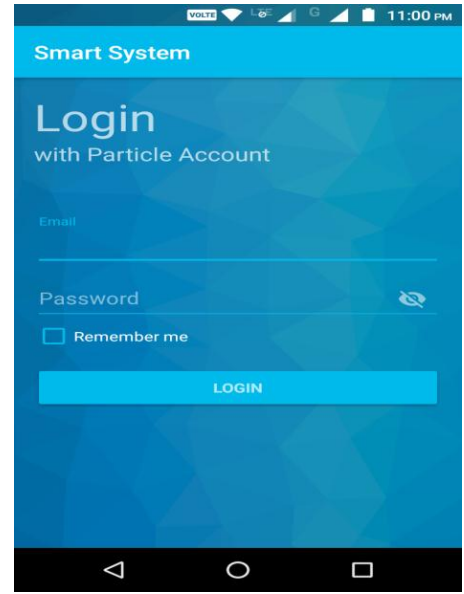

*Figure 5: Mobile Application for APCS*

### 3.3 Thingspeak

Thingspeak is a web-based open API IoT information platform that can store the sensor data of a wide variety of IoT applications [9]. It is also used to combine different varieties of sensor data for analysis and thus helps the user in making the right decisions. The data from Thingspeak can be outputted in the graphical format at the web level. Internet helps in the communication of Thingspeak and other devices. It could analyze, retrieve, save/store, observe and work on the sensed data from the devices such as Raspberry-pi, Arduino, particle photon, Intel Galileo etc., to the sensors.

Thingspeak helps in a sensor based logging applications, social networking of objects/ things with updated status and location tracing applications. Alternatively, It can also be used in home automation products that were connected to the internet. The primary feature of the Thingspeak functionality is the term "channel" that have various fields for defining the status of varied sensed data, location and data. The data from the Things speak can be processed and analyzed only after creating the channels. The data that is stored in the Thingspeak can be utilized for visualization purpose using MATLAB and can respond to the data with tweets and other forms of alerts also. It also provides a feature to create a public based channel to analyze and estimate it through the public. All these activities mentioned above are features of a Cloud and hence Thingspeak is treated as a cloud.

The IoT Helps to bring all things together and permits us to communicate with our very own things and even more curiously allows objects/things to interact with other 'things'.

## IV. SYSTEM DESIGN AND IMPLEMENTATION

The implementation of the proposed system (APCS) is shown in Figure 6. It shows the connectivity of various devices to particle photon chip. IR and Relay devices get power from GND and +5V pins from Particle device. The IR sensor is usually used for obstacle detection but in APCS it is used for detecting the direction of motion. It shows how the two IR sensors are placed at the entrance of the room. The phase difference between the readings of the two sensors can be used to detect the directions of the movement. The particle photon is continuously reading both the sensors values. The person while moving in or out of the room will pass through these sensors one by one. If the IR1 sensor, gives +5V before the IR2 sensor then the person is moving IN the room. If IR2 sensor gives +5V before the IR1 sensor then the person is moving OUT of the room. Whenever a person is detected entering or leaving the room the count is incremented or decremented respectively. Thus we can know the number of persons available inside the room at any moment and can check whether the room is empty or occupied. Accordingly, APCS switch on or off the electrical appliances.

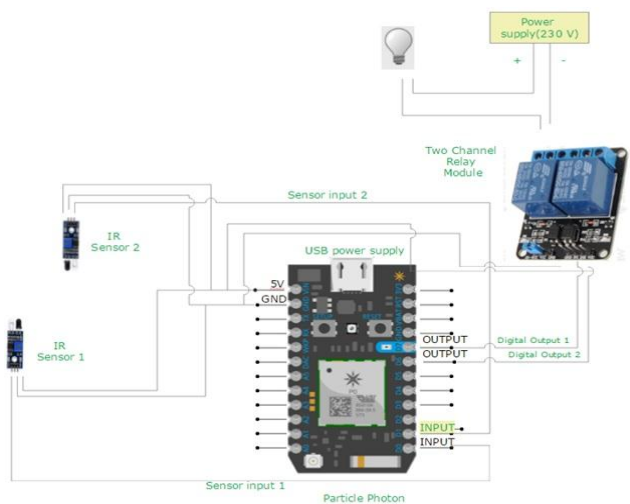

*Figure 6: Implementation of Automated Power Conservation System*

## V. ALGORITHM AND FLOWCHART FOR APCS

The algorithmic approach for APCS was given below for better understanding of the user. The Particle.publish() method uploads count data to the Thingspeak cloud for every event occurs.

| Algorithm 1: An Automated Power Conservation System |
|---|
| **Input:** Two IR sensor inputs IR1 and IR2 |
| **Output:** Outputs light and fan via Relay module |
| **Initialization:** *count* ← *0* |
| 1.   Loop |
| 2.       if *IR1 == HIGH* then |
| 3.       while 15sec completes do |
| 4.           if *IR2==HIGH* then |
| 5.        *count++* |
| 6.    *particle. publish(count) ;* |
| 7.    end |
| 8.     end |
| 9.    end |
| 10.   if *IR2==HIGH* then |
| 11.   while *15sec completes* do |
| 12.   if *IR1==HIGH* then |
| 13.   *count -- ;* |
| 14.   *particle. publish(count) ;* |
| 15.   end |
| 16.   end |
| 17.   end |
| 18.   if *count <= 0* then |
| 19.   light←LOW; |
| 20.   fan←LOW; |
| 21.   else |
| 22.   light←HIGH; |
| 23.   fan←HIGH; |
| 24.   end |
| 25.   EndLoop |

### 4.1 Flow Chart

Figure 7 gives the flowchart of APCS while it is operating in Automatic mode. Initially, both the sensors are waiting for obstacle detection if the IR1 sensor detects signal first then 15sec after the IR2 sensor detects signal. It means person entered the room so count variable increments and count upload into the Thingspeak. if the count equals to zero the microcontroller turns off the lights and fans. If the IR2 detects first then IR1 means person left the room so count decrements. Every action, the particle will upload data into the cloud.

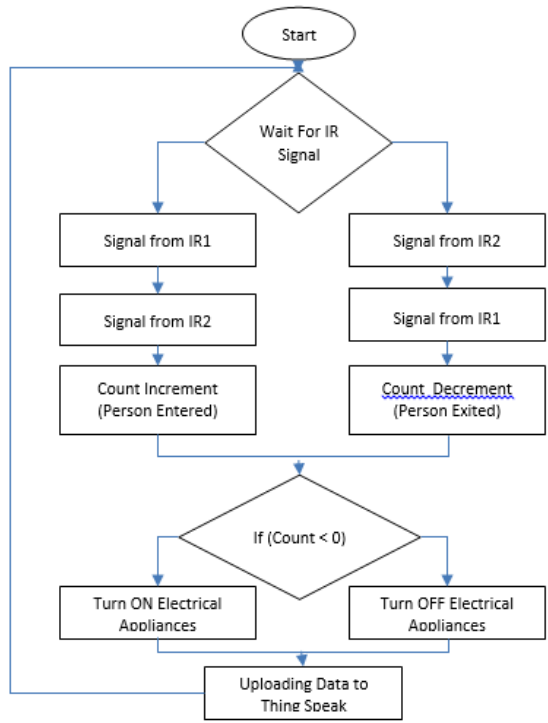

*Figure 7: Flowchart of APCS while operating in an Automatic mode*

The particle photon continuously reading values from sensors and uploading into Thingspeak using webhook which is already configured in particle web. The mobile application gets these data from Thingspeak using JSON objects.

## VI. EXPERIMENTAL SETUP

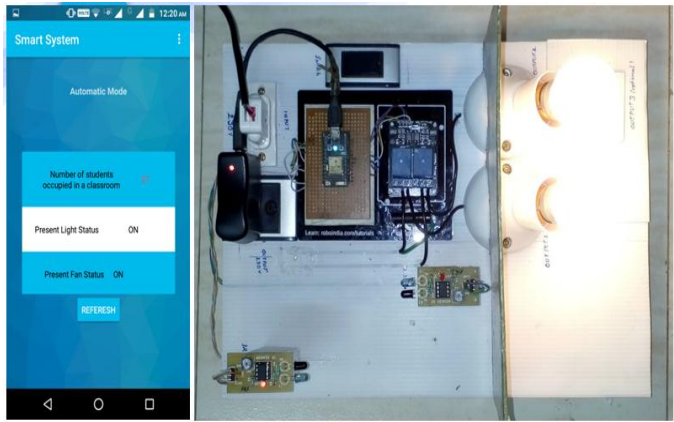

*Figure 8: APCS Experimental Setup*

Experimental setup for the proposed system (APCS) is shown in Figure 8. APCS program is flashed into particle photon. APCS android app has been installed on the smartphone. The user gets all the information about the devices with the help of the APCS mobile app. The user interacts with the system by simply logging into the app with help of the unique user id and password.

## VII. EXPERIMENTAL RESULTS

The proposed system APCS has been tested successfully for both manual and automatic operation of electrical appliances based on the presence of the persons in the classroom. APCS can be treated as a novel approach to conserving power and it is calculated that almost 15% of power can be saved per month. Conserving power has a direct impact on money savings also. Even though the APCS is developed keeping in view of the university, it is also proved that it is best-suited home automation and industrial automation where one can conserve more power from being wasted.

## VIII. CONCLUSION AND FUTURE WORK

This paper proposes a low cost, secure, ubiquitously accessible, remotely controlled solution. The approach discussed in the paper is novel and has achieved the target to control electrical appliances remotely using the Wi-Fi technology to connect system parts, satisfying user needs and requirements. Looking at the current scenario we have chosen the Android platform so that most of the people can get the benefit. The technology is easy to use and can benefit the naive people that have no technical background. The proposed system is better from the scalability and flexibility point of view than the commercially available automation systems.

In future proposals can be made to build a cross-platform system that can be deployed on various platforms like iOS, Windows etc. This System can be extended to a number of electrical appliances. INPUTS and OUTPUTS can be extended by using Basic Logic gates. Many other devices like Security cameras can be controlled, allowing the user to observe activity around a room. Security systems can include motion sensors that will detect any kind of unauthorized movement and notify the user. The scope of this project can be expanded to the entire organization but not for a single room.

**Authors Profile**

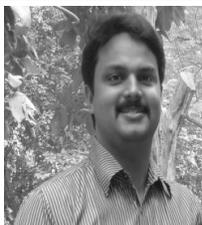

Chandra Sekhar Sanaboina is presently pursuing Ph.D. under the guidance of Prof. Pallamsetty Sanaboina from Andhra University. He obtained his B.Tech. In Electronics and Computer Science Engineering in 2005 and M.Tech in Computer Science and Engineering from Vellore Institute of Technology in 2008. He has over 10 years of teaching experience and currently working as Assistant Professor in the Department of Computer science and Engineering, JNTUK Kakinada. His areas of interests include Wireless Sensor Networks, Internet of Things, Machine Learning and Artificial Intelligence.

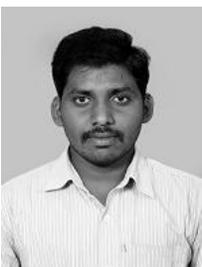

Harish Bommidi completed his M.Tech in Computer Science and Engineering from UECK, JNTUK Kakinada in the year 2017. He has completed his B.Tech in Computer Science and Engineering, from Vignan Institute of Information Technology, Duvvada, Visakhapatnam. He has done his M.Tech Project under the guidance of Chandra Sekhar Sanaboina, Assistant Professor in the Department of Computer Science and Engineering, JNTUK Kakinada. Presently he is working for TCS, Bangalore as Assistant Systems Engineer. His Areas of interests include Wireless SensorNetworks, Internet of Things.